\documentclass[a4paper,12pt]{article}

\usepackage[T2A]{fontenc}
\usepackage[utf8]{inputenc}   
\usepackage[russian]{babel}

\usepackage{CJKutf8}

\usepackage{amsmath,amstext,amsfonts,amsthm,amssymb}
\usepackage{eucal}
\usepackage{graphicx}
\usepackage{tikz-cd}
\usetikzlibrary{patterns}
\usetikzlibrary{calc}

\renewcommand{\subsubsection}[1]{\addtocounter{subsubsection}{1}
{\ \\[3pt]\bf \thesubsubsection. \  #1} }

\swapnumbers

{  \theoremstyle{definition}

}
\newcommand{\Mat}{\operatorname{Mat}}

\newcommand{\isom}{\overset{\sim}{=}}
\newcommand{\lra}{\longrightarrow}

\newcommand{\bea}{\begin{eqnarray*}}
\newcommand{\eea}{\end{eqnarray*}}
\newcommand{\bean}{\begin{eqnarray}}
\newcommand{\eean}{\end{eqnarray}}

\newcommand{\fg}{\mathfrak g}
\newcommand{\fh}{\mathfrak h}

\newcommand{\fosp}{\mathfrak{osp}}

\newcommand{\fsl}{\mathfrak{sl}}
\newcommand{\fso}{\mathfrak{so}}

\newcommand{\fsp}{\mathfrak{sp}}

\newcommand{\CG}{\mathcal{G}}

\newcommand{\BC}{\mathbb{C}}

\newcommand{\BQ}{\mathbb{Q}}
\newcommand{\BR}{\mathbb{R}}
\newcommand{\BZ}{\mathbb{Z}}

\newcommand{\nc}{\newcommand}

\nc{\Id}{\text{Id}}
\nc{\la}{\lambda}

\begin{document}





\centerline{\bf MOMENT MAP AND MATRIX INTEGRALS}

\

\centerline{\it Gaussian separation of variables}

\bigskip\bigskip

\centerline{Vadim Schechtman}


\begin{CJK}{UTF8}{min}


\end{CJK}

\bigskip\bigskip

\centerline{April 22, 2021}


\bigskip\bigskip

 

\

\centerline{\bf Abstract}

\

We discuss the geometry behind some integrals related to structure constants of the Liouville conformal field theory.

\

\

\centerline{\bf \S 0. Introduction. Three incarnations of a quadratic map}

\

This note is a followup of [BS]; it is mostly a review of known results.  
We discuss some geometry lying behind the computations from [ZZ] and [BR], and their $p$-adic and adelic analogs.

\ 

{\bf 0.1. Moment ternoon.} Let $K$ be a field of characteristic $\neq 2$. We will discuss certain quadratic map between two affine spaces
$$
\mu:\ K^6\lra K^3
\eqno{(0.1.1)}
$$
It may be introduced in three ways.

(a) As an {\it exterior multiplication}
$$
\mu:\ K^3\times K^3\lra \Lambda^2(K^3),\ (x, y)\mapsto x\wedge y.
\eqno{(0.1a)}
$$

(b) As a {\it moment map}. Regard $X(K) = K^6$ as the cotangent space to $Y(K) = K^3$; the group 
$H = SO_3(K)$ acts on $Y(K)$ in an obvious way; this action is Hamiltonian, and  $\mu$ is the momentum map, 
 $K^3$ on the right being identified with $\fh^* := Lie(H)^*$:
$$
\mu:\ T^*Y \lra \fh^*
\eqno{(0.1b)}
$$ 

This is the archetypical moment map, wherefrom its very name has appeared.
Its three components are "angular momenta"\ .

(c) As a {\it quotient map}. Identify $X(K)$ with the space of $2\times 3$ matrices; the group 
$G = SL_2(K)$ acts upon $X(K)$ from the left, and $\mu$ may be identified (at least birationally) 
with the quotient map
$$
\mu:\ X(K) \lra  G(K)\backslash X(K) = Y(K),
\eqno{(0.1c)}
$$
$Y$ being identified with a categorical quotient of $X$, by the Igusa criterion, cf. [I]. 

\

{\bf 0.2.} In [BR] the map $\mu$ (for $K = \BR$) has been used for a computation 
of certain triple integral $I(a, b, c; \BR)$, $a, b, c\in \BC$ over $Y(\BR)$, see (1.1.7) below.

A similar integral for $K = \BC$ has appeared previously in [ZZ] (cf. also [Z]). We can take $K$ to be a nonarchimedian local field; the integral $I(a, b, c; \BQ_p)$
has been introduced and  computed in [BS]. In {\it op. cit.} a $q$-deformation of $I(a,b,c)$ is discussed 
as well.

The upshot of the trick from [BR] is that an integral $I(a, b, c; K)$ over $Y(K)$ 
is represented as a ratio of two Gaussian integrals over $X(K)$ and $Y(K)$.  

From our viewpoint it might be considered as an integral over a fiber  
$X_t := \mu^{-1}(t),\ t\in Y(K)$, and indeed, in the original definition in [ZZ] $I(a, b, c; \BC)$ has appeared as an integral over $G(\BC)$, see \S 1 below.

{\bf 0.3.} 
I am much obliged to M.Finkelberg for consultations; among others things he explained to me that 
0.1 is a particular case (and a part) of a general superalgebra construction described e.g. in [BFT], see \S 4 below.   


\

\


\centerline{\bf \S 1. Some geometry behind an integral}

\

{\bf 1.1. Complex and real integrals.} The following integral appears in [ZZ] (4.17)
$$
I(\sigma_1,\sigma_2,\sigma_3;\BC) = \int_{\BC^3}\prod_{i\in\BZ/3\BZ}(1 + |z_i|^2)^{-2\sigma_i}
\prod_{i\in\BZ/3\BZ}(z_i - z_{i+1})^{-2 -  2\nu_{i+2}}
\prod_{i\in\BZ/3\BZ} |d^2z_i|,
\eqno{(1.1.1)}
$$
where $|d^2z| = dx dy,\ z = x + iy$. Here 
$$
\nu_i := \sigma_i - \sigma_{i+1} - \sigma_{i+2};
\eqno{(1.1.2)}
$$
thus
$$
- 2\sigma_i = \nu_i + \nu_{i+1}.
\eqno{(1.1.3)}
$$
Its value is  
$$
I(\sigma_1,\sigma_2,\sigma_3;\BC) = \pi^3\Gamma(\sum_i\sigma_i  - 1)
\frac{\prod_{i\in\BZ/3\BZ}
\Gamma(-\nu_i)}{\prod_{i\in\BZ/3\BZ}\Gamma(2\sigma_i)}
\eqno{(1.1.4)}
$$
A real version of (1.1.1) looks as follows:
$$
I(\sigma_1,\sigma_2,\sigma_3;\BR) = \int_{\BR^3}\prod_{i\in\BZ/3\BZ}(1 + x_i^2)^{-2\sigma_i}
\prod_{i\in\BZ/3\BZ}|x_i - x_{i+1}|^{- 2(1 + \nu_i)} dx_1dx_2dx_3,
\eqno{(1.1.5)}
$$
cf. [BS] (a), \S 6. After a change of variables
$$
x_i = \tan \alpha_i
$$
it becomes 
$$
I(\sigma_1,\sigma_2,\sigma_3;\BR) = \frac{1}{8}\int_{[-\pi, \pi]^3}
\prod_{i\in\BZ/3\BZ}|\sin(\alpha_i - \alpha_{i+1})|^{- 2(1 + \nu_i)} d\alpha_1d\alpha_2d\alpha_3,
\eqno{(1.1.6)}
$$
in this form it appears in [BR], \S 5,  (19). The computation from {\it loc. cit.} gives 
the following answer:
$$
I(\sigma_1,\sigma_2,\sigma_3;\BR) = \pi^{3/2}\Gamma(\sum_i\sigma_i - 2)
\frac{\prod_i \Gamma(-1/2 - \nu_i)}
{\prod_i\Gamma(2\sigma_i -1)}
\eqno{(1.1.7)}
$$



\

{\bf 1.2. Reduction to Gaussian integrals.} 
The authors of [BR] have proposed an elegant method for computing $I(\sigma_1,\sigma_2,\sigma_3;\BR)$. 

Consider the real vector space $V = \BR^3$ and the multpilication map
$$
\mu:\ V\times V\lra \Lambda^2V,\ (x,y) \mapsto x\wedge y
\eqno{(1.2.1)}
$$
If we identify the $6$-dimensional vector space $V\times V$ with the space of $2\times 3$ matrices 
and the $3$-dimensional vector space $\Lambda^2V$ with $\BR^3$ using a base 
$\{e_i\wedge e_{i+1}\}$  
then $\mu$ will be a "moment"\  map 
$$
\mu:\ X = M_{2,3} \lra Y = \BR^3
\eqno{(1.2.3)}
$$
which assigns to a matrix $A$ its three $2\times 2$ minors, 
$$
\mu(A) = (|A_{12}|, |A_{23}|, |A_{31}|).
$$

\

{\it Structure of the formula}

\

Let us consider the formula (1.1.7). Denote
$$
A(\BR) = \pi^{3/2}\Gamma(\sigma_1 + \sigma_2 + \sigma_{3} - 2),
$$
$$
B(\BR) = \prod_{i=1}^3 \Gamma(-1/2 - \nu_i),
$$
and
$$
C(\BR) = \prod_{i=1}^3 \Gamma(2\sigma_i -1).
$$
Then
$$
I(\BR) = A(\BR)\cdot\frac{B(\BR)}{C(\BR)}.
\eqno{(1.2.4)}
$$
Bernstein and Reznikov interpret the numerator $B(\BR)$ (resp. the denominator $C(\BR)$)  as 
a Gaussian integral over $X$ (resp. $Y$); the prefactor $A(\BR)$ appears due, roughly speaking, to the 
map $\mu$.

A complex version of the above computation is described in [BS] (b). 

\

{\bf 1.3. Fibers of the moment map and $SL_2$.} Let us try to understand the above computation "motivically"\ . Consider various fibers
$$
X_t := \mu^{-1}(t)\subset X,\ t\in Y;
$$
they are subvarities of $X = M_{2,3}(F)$ where $F = \BC, \BR, \BQ_p, \ldots$ .

Among them there is a distinguished  one
$$
Z = X_0\subset X = M_{2,3}, 
$$
- the subvariety of matrices of rank $1$, $\dim Z = 4$. For $t\neq 0$ $\dim X_t = 3$. 

We may say that morally the "motive"\ $[X_y]$ is a ratio 
$$
[X_t] = \frac{[X]}{[Y]}
\eqno{(1.3.1)}
$$
and ask if our integral $I$ may be interpreted as an integral over  $X_t$. 

{\bf 1.3.1.} We remark that the group $G = SL_2(F)$ is acting upon $X$ by left 
multiplication and respects the fibers $X_t$, due to the equality of minors
$$
(gA)_{ij} = gA_{ij},\ g\in G,\ A\in X.
$$
If we pick a point $x\in X_t$, the map
$$
\nu_x:\ G \lra X_t,\ g\mapsto gx
$$
is a birational isomorphism since both guys are $3$-dimensional. 

Therefore we may guess that probably our integral $I$ may be interpreted as an integral over $G$. 

\

{\bf 1.3.2.} Note that there is also an action of another group $H = SO(3)$ on $X$, along the rows 
of a matrix. 

We can identify $X$ with $T^*V$, and then $\mu$ will be the moment map for this 
action, if we identify $Y$ with $\fh = Lie(H)$, see [A], Appendix 5. 

The fibers $X_t\subset X$ will be Lagrangian 
subvarieties.  

\

{\bf 1.4.} The above guess turns out to be true. Consider the case $F = \BC$ treated in [ZZ].   
Originally $I(\BC)$ is defined in [ZZ] as follows:
$$
I(\sigma_1,\sigma_2,\sigma_3;\BC) = 
$$
$$
\int_{G(\BC)} (|b|^2 + |d|^2)^{-2\sigma_1} 
(|a + b|^2 + |c + d|^2)^{-2\sigma_2}(|a|^2 + |c|^2)^{-2\sigma_3}d\mu_{G(\BC)},\ 
\eqno{(1.4.1)}
$$


where 
$$
g = \left(\begin{matrix} a & b\\c & d
\end{matrix}\right) \in G(\BC),
$$ 
and 
$$
d\mu_G(g) = 4d^2 da\ db\ dc\ dd  
$$ 
is a Haar measure on $G(\BC)$, see {\it op. cit.} (4.14), (2.44).  

The integral (1.4.1) takes a form (1.1.1) after a change of variables 
$$
z_1 = b/d, z_2 = (a + b)/(c + d), z_3 = a/c,
\eqno{(1.4.2)}
$$
cf. {\it op. cit.}  (4.15). 

It comes out in turn from an integral
$$
J(z_1, z_2, z_3) := \pi^3\int_{G(\BC)} \prod_{i=1}^3 |g\cdot z_i|^{2\sigma_i} d\mu_G(g) = 
$$
$$
= \prod_{i\in\BZ/3\BZ} 
|z_i - z_{i+1}|^{2\nu_{i+2}} 
I(\sigma_1, \sigma_2, \sigma_3;\BC) 
\eqno{(1.4.3)}
$$
where
$$
|g\cdot z|^2 := |az + b|^2 + |cz + d|^2,
$$
cf. {\it op. cit.} (4.13). 
This integral is a "quasiclassical limit"\ of a three-point correlation function 
in the Liouville CFT. 

\

\centerline{\bf \S 2. $p$-adic and adelic}

\

{\bf 2.1. $p$-adic.} Consider the $p$-adic field $K = \BQ_p$; for $x\in \BQ_p$ we set 
$$
|x|_p := p^{-v_p(x)};
$$
let $d_px$ denote the Haar measure on $K$ normalised in such a way that
$$
\int_{\BZ_p}d_px = 1.
$$
Define a function $\psi_p:\ K\lra \BR$ by 
$$
\psi_p(x) = \max\{|x|_p, 1\},
\eqno{(2.1.1)}
$$
- it is a $p$-adic analog of $|z|^2 + 1,\ z\in \BC,$ cf. [BS], 2.5.

A $p$-adic version of (1.1.1) is
$$
I(\sigma_1,\sigma_2,\sigma_3;\BQ_p) = 
$$
$$
\int_{\BQ_p^3}\prod_{i\in\BZ/3\BZ} \psi_p(x)^{-2\sigma_i}
\prod_{i\in\BZ/3\BZ}|x_i - x_{i+1}|_p^{-1 -  \nu_{i+2}}
d_px_1d_px_2d_px_3, 
\eqno{(2.1.2)}
$$ 
cf. [BS] (a), 1.8. Its value has been computed in [B], [BS] (a), Theorem 1.9.

Define a function
$$
\Gamma_p(s) := \frac{1}{1 - p^{-s}},\ s\in\BC
\eqno{(2.1.3)}
$$
Then
$$
I(\sigma_1,\sigma_2,\sigma_3;\BQ_p) = \Gamma_p(2)^{-1}\Gamma_p(\sum_i\sigma_i  - 1)
\frac{\prod_{i\in\BZ/3\BZ}
\Gamma_p(-\nu_i)}{\prod_{i\in\BZ/3\BZ}\Gamma_p(2\sigma_i)}.
\eqno{(2.1.4)}
$$
Note that this result contains some regularisation behind the scene. Namely, the integral (2.1.2) 
is written as a sum of integrals which converge in different half-spaces of values for parameters 
$\sigma_i$; each of these summands is calculated, and is in the obvious way extended to a meromorphic function on $\BC^3$. Our functions are similar to the "$p$-adic beta function"\ from [GGPS], Ch. II, 5.5. 

\

{\it A matrix version}

\

The same integral is equal to a matrix one 
$$
I(\sigma_1,\sigma_2,\sigma_3;\BQ_p) = 
$$
$$
\int_{G(\BQ_p)} (|b|_p^2 + |d|_p^2)^{-2\sigma_1} 
(|a + b|_p^2 + |c + d|_p^2)^{-2\sigma_2}(|a|_p^2 + |c|_p^2)^{-2\sigma_3}d\mu_{G(\BQ_p)},\ 
\eqno{(2.1.5)}
$$

{\bf 2.2. Balance.} In {\it op. cit.} a slightly different $p$-adic Gamma is used, namely
$$
\Gamma_p(s) := \frac{1 - p^{-1}}{1 - p^{-s}}.
$$
It appears naturally in calculation but goes from the final answer since in it we have four Gammas both 
in the numerator and in the denominator. 

This circumstance is lucky for the Euler product (see below).

\

{\bf 2.3. Global: an Euler product.}
Let $P$ denote the set of all rational  primes $p$.
Consider a product 
$$
I(\sigma_1,\sigma_2,\sigma_3)_A := \prod_p \Gamma_p(2)^{-1}\Gamma_p(\sum_i\sigma_i  - 1)
\frac{\prod_{i\in\BZ/3\BZ}
\Gamma_p(-\nu_i)}{\prod_{i\in\BZ/3\BZ}\Gamma_p(2\sigma_i)}
\eqno{(2.3.1)} 
$$
This product converges for $(\sigma_1, \sigma_2, \sigma_3)\in\BC^3$ such that
$$
D: \Re(\sigma_i) > 1/2,\ \sum_i \Re(\sigma_i) > 1,\ 
\Re(\sigma_i) + \Re(\sigma_{i+1}) - \Re(\sigma_{i+2}) > 1
\eqno{(2.3.2)} 
$$
This domain is nonempty: it contains a subset 
$$
\Re(\sigma_1) = \Re(\sigma_2) = \Re(\sigma_3) > 1.
$$
The value obviously is
$$
I(\sigma_1,\sigma_2,\sigma_3)_A  = 
\zeta(2)^{-1}\zeta_(\sum_i\sigma_i  - 1)
\frac{\prod_{i\in\BZ/3\BZ}
\zeta(-\nu_i)}{\prod_{i\in\BZ/3\BZ}\zeta(2\sigma_i)}
\eqno{(2.3.3)}
$$

Let $A^f(\BQ)$ be the ring of finite ad\`eles for $\BQ$.
It is tempting to conjecture that $I(\sigma_1,\sigma_2,\sigma_3)_A$ is equal 
to some integral over $A^f(\BQ)$. 

The adelic integral similar to (1.4.1), (2.1.2) does not make sense: one needs some regularization.
One of the possible ways would be a ratio of two Gaussian integrals. 

\

\centerline{\bf \S 3. $q$-deformations}

\

{\bf 3.1.} A $q$-deformation of the ZZ integral is proposed and calculated in [BR] (a), where it is formulated 
in the form of an "exotic"\ Macdonald constant term identity, certain generalization 
of this identity for the root system $A_2$. In this form it was discovered earlier by W.Morris, 
[M].

$q$-deformed Liouvelle triple correlators, in their gauge theory avatars, appear in the physical papers, 
cf. [CPT] and references therein. 

It is not excluded that this identity may be proven by BR Gaussian trick as well, with Jackson integrals 
replacing the usual ones.

\

\centerline{\bf \S 4. Second moment and supersymmetry}

\

{\bf 4.1. Orthosymplectic Lie superalgebra.} The action of the groups $G = SL_2$ and $H = SO(3)$ 
on the space $X = \Mat_{2,3}$ commute with each other, in other words, the group
$$
\CG = G\times H
$$
is acting on $X$.  The groups $G$ and $H$ look similar if one recalls that $SL_2 = Sp(2)$; 
both of them have type $B_1 = C_1 = A_1$.  

Consider a vector superspace $V = V_0 \oplus V_1$, $\dim V_0 = 3,\ \dim V_1 = 2$ equipped with a nondegenerate 
bilinear form $B$ which is symmetric on $V_0$ and skew-symmetric on $V_1$, such that 
$V_0$ and $V_1$ are orthogonal with respect to $B$.  

The Lie superalgebra $\frak{osp}(3|2)$ is by definition a subalgebra of 
$\frak{gl}(3|2)$ consisting of endomorphisms respecting $B$. 

If $B$ is given by the matrix
$$
B = \left(\begin{matrix}1 & 0 & 0 & 0 & 0\\
0 & 0 & 1 & 0 & 0\\
0 & 1 & 0 & 0 & 0\\
0 & 0 & 0 & 0 & 1\\
0 & 0 & 0 & -1 & 0
\end{matrix}\right)
$$
then $\fosp(3|2)$ consists of matrices of the form
$$
\left(\begin{matrix}
0 & -u & -v & x & x'\\
v & a & 0   & y & y'\\
u & 0 & -a & z & z'\\
-x' & -z' & -y' & d & e\\
x & z & y & f & -d
\end{matrix}\right),
$$
cf. [Mu], 2.3. 

Thus
$$
\fg_0 = \fso(3)\oplus \fsp(2) =  \fso(3)\oplus \fsl_2,
$$
so 
$$
\dim\fg_0 = \dim\fg_1 = 6.
$$
We have 
$$
\fg_0 = Lie(\CG),
$$
and on the other hand one can identify 
$$
\fg_1\isom X
$$
so that the action of $\CG$ on $X$ will be the adjoint action.

\

{\bf 4.2. The moment ternoon.}
We have a superbracket map
$$
\mu_{(a)}:\ \fg_1 \lra \fg_0 = Lie(G)\oplus Lie(H),\  \mu_{(a)}(x) = \frac{1}{2}[x,x]
\eqno{(4.2.1a)}
$$
whose first component
$$
\mu_{(a1)}:\ \fg_1 \lra Lie(G) 
$$
is the map $(0.1a)$, at the same time we have a partner
$$
\mu_{(a2)}:\ \fg_1 \lra Lie(H).
$$

Next, we have the moment map
$$
\mu_{(b)}:\ \fg_1 \lra \fg_0^* = Lie(G)^*\oplus Lie(H)^*,\ 
\eqno{(4.2.1b)} 
$$  
whose first component
$$
\mu_{(b1)}:\ \fg_1 \lra Lie(G)^* 
$$
is the map $(0.1b)$, 

and finally we have the two quotient maps
$$
\mu_{(c1)}:\ \fg_1 \lra \fg_1/G
\eqno{(4.2.1c1)}
$$
which is the map $(0.1c)$, and its partner
$$
\mu_{(c2)}:\ \fg_1 \lra \fg_1/H.
\eqno{(4.2.1c2)}
$$

{\bf 4.3.} For a generalization to an arbitrary $\fosp(m|2n)$ see e.g. [BFT], 2.1, 2.8.

\bigskip\bigskip

\centerline{\bf References}

\bigskip\bigskip

[A] V.Arnold, Mathematical methods in classical mechanics

[BR] J.Bernstein, A.Reznikov, Estimates of automorphic functions, {\it Mosc. Math. J.} {\bf 4} (2004), 19-37.

[BFT] A.Braverman, M.Finkelberg, R.Travkin, Orthosymplectic Satake equivalence, arXiv:1912.01930.

[B] Bui Van Binh, Complex and $p$-adic Selberg integrals, and Dyson-Macdonald identities, PhD Thesis, Toulouse, 2014, 
\newline https://www.math.univ-toulouse.fr/~schechtman/buivanbih-these.pdf

[BS] Bui Van Binh, V.Schechtman, (a) Remarks on a triple integral, {\it Mosc. Math. J.} {\bf 13} (2013), no. 4, 585–600; (b) Invariant functionals and 
Zamolodchikovs' integral, {\it Funct. Anal. Appl.} {\bf 49} (2015), no. 1, 71–74.

[CPT] I.Coman, E.Pomoni, J.Teschner, Trinion comformal blocks from topological strings, arXiv:1906.06351.


[GGPS] I.M.Gelfand, M.I.Graev, I.I.Pyatetsky-Shapiro, Representation theory and automorphic functions, {\it  Generalized functions} {\bf 5}. 

[I] J-I.Igusa, Geometry of absolute irreducible representations, In: {\it Number theory, algebraic geometry 
and commutative algebra}, {\bf 1973}. 

[M] W.G.Morris II, Constant term identities for finite and affine root systems: conjectures and theorems, 
PhD Thesis, University of Wisconsin - Madison, 1982. 

[Mu] I.Musson, Lie superalgebras and enveloping algebras


[Z] Al.Zamolodchikov, On the three-point function in minimal Liouville gravity, arXiv:hep-th/0505063.

[ZZ] A.Zamolodchikov, Al.Zamolodchikov, Conformal bootstrap in Liouville field theory, 
{\it Nucl. Phys.} {\bf B 477} (1996), 577-605.

\newpage



\end{document}